\documentclass[twocolumn]{aastex631}

\usepackage{amsmath}
\usepackage{graphicx}
\usepackage{hyperref}

\newcommand{\oiii}{[O\,{\sc iii}]}

\begin{document}

\title{Stacking X-ray Observations of ``Little Red Dots'': Implications for their AGN Properties}

\correspondingauthor{Minghao Yue}
\email{myue@mit.edu}

\author[0000-0002-5367-8021]{Minghao Yue}
\affiliation{MIT Kavli Institute for Astrophysics and Space Research, 77 Massachusetts Ave., Cambridge, MA 02139, USA}

\author[0000-0003-2895-6218]{Anna-Christina Eilers}
\affiliation{MIT Kavli Institute for Astrophysics and Space Research, 77 Massachusetts Ave., Cambridge, MA 02139, USA}

\author[0000-0001-8211-3807]{Tonima Tasnim Annana}
\affiliation{Department of Physics and Astronomy, Wayne State University, 666 W Hancock St, Detroit, MI 48201, USA}

\author{Christos Panagiotou}
\affiliation{MIT Kavli Institute for Astrophysics and Space Research, 77 Massachusetts Ave., Cambridge, MA 02139, USA}

\author[0000-0003-0172-0854]{Erin Kara}
\affiliation{MIT Kavli Institute for Astrophysics and Space Research, 77 Massachusetts Ave., Cambridge, MA 02139, USA}

\author[0000-0002-7562-485X]{Takamitsu Miyaji}
\affiliation{Instituto de Astronom\'ia, Universidad Nacional Aut\'onoma
 de M\'exico Campus Ensenada, A.P. 106, Ensenada, BC 22800, Mexico}



\begin{abstract}
Recent {\em James Webb Space Telescope (JWST)} observations have revealed a population of compact extragalactic objects at $z\gtrsim4$ with red near-infrared colors, which have been dubbed as ``Little Red Dots" (LRDs). The spectroscopically-selected LRDs exhibit broad H$\alpha$ emission lines, which likely indicates that type-I active galactic nuclei (AGN) are harbored in the galaxies' dust-reddened cores. However, other mechanisms, like strong outflowing winds, could also produce broad H$\alpha$ emission lines, and thus, the nature of LRDs is still under debate. We test the AGN hypothesis for LRDs by stacking the archival {\em Chandra} observations of 34 spectroscopically-selected LRDs. 
We obtain tentative detections in the soft $(0.5-2\text{ keV})$ and hard $(2-8\text{ keV})$ X-ray bands with $2.9\sigma$ and $3.2\sigma$ significance, and with $4.1\sigma$ significance when combining the two bands. 
Nevertheless, we find that the soft (hard) band $3\sigma$ upper limit is $\sim1$dex ($\sim 0.3$dex) lower than the expected level from the $L_\text{X}-L_{\text{H}\alpha}$ relation for typical type-I AGNs. Our results indicate that AGN activity is indeed likely present in LRDs, though these objects have significantly different properties compared to previously identified type-I AGNs, \textcolor{black}{i.e., 
LRDs may have intrinsically weak X-ray emissions.
We find it difficult to explain the low $L_\text{X}/L_{\text{H}\alpha}$ ratios observed in LRDs solely by absorption. It is also unlikely that fast outflows have major contributions to the broad H$\alpha$ lines.}Ther
Our findings indicate that empirical relations (e.g., for black hole mass measurements) established for typical type-I AGNs should be used with caution when analyzing the properties of LRDs.
\end{abstract}



\section{Introduction} \label{sec:intro}

%Active Galactic Nuclei (AGNs) are supermassive black holes (SMBHs) that are actively accreting materials, releasing copious amounts of energy that can have fundamental impacts on their host galaxies and environments (ref). 
%A comprehensive understanding of the AGN population and their properties is a crucial task in extragalactic astronomy.

The recent launch of the {\em James Webb Space Telescope} (JWST) has opened new windows towards studies of distant galaxies and supermassive black holes (SMBHs).
One particularly exciting discovery by {\em JWST} is the abundant population of so-called ``little red dots" (LRDs). LRDs are compact objects at $z\gtrsim4$ with very red near-infrared color, which have been found by many {\em JWST} survey programs \citep[e.g.,][]{labbe23, kocevski23, harikane23, matthee24, greene23}. 
Their spectra exhibit broad H$\alpha$ emission lines, suggesting type-I active galactic nuclei (AGN) activity in their cores.
%usually with luminosities of $L_{\text{H}\alpha}\gtrsim10^{42}\text{erg s}^{-1}$ and full-width half maxima (FWHM) of $\text{FWHM}_{\text{H}\alpha}\gtrsim1000\text{ km s}^{-1}$. 
%The large H$\alpha$ line widths indicate that LRDs might be a new population of dust-reddened type-I active galactic nuclei (AGNs), with implied SMBH masses of $M_\text{BH}\sim10^6-10^7M_\odot$. 
The number density of LRDs at $z\sim5$ is about $10^{-5}-10^{-4}\text{ Mpc}^{-3}\text{mag}^{-1}$ \citep[e.g.,][]{greene23}, which is \textcolor{black}{close to the entire X-ray AGN population at similar redshifts and is} $\sim1-2$ dex higher than the faint end of quasar luminosity functions \citep[e.g.,][]{matsuoka18,jts23}. This comparison suggests that LRDs might be a population of previously unexplored dust-reddened AGNs, and that our current picture of SMBH-galaxy co-evolution might be incomplete.% Correctly characterizing these objects is thus crucial to the understanding of SMBH-galaxy coevolution and cosmic reionization.

%The discovery of LRDs indicates that the current picture of AGNs might be highly incomplete. %
However, the properties of LRDs are still largely unclear, and a few models have been proposed to explain the observed features of these objects. \citet{kocevski23} found that the spectral energy distributions (SEDs) of LRDs can be explained either by a dust-obscured AGN plus a blue galaxy or by an unobscured AGN combined with a red galaxy. \citet{labbe23} and \citet{greene23}  suggested that LRDs might be heavily obscured AGNs and that the broad emission lines and rest-frame UV continuum of LRDs might arise from the scattered light of the central AGNs.
\cite{matthee24} proposed that LRDs represent a transition phase from heavily obscured AGNs to unobscured quasars. This model is in line with \citet{nobo23}, who found that LRDs share similar features with blue-excess dust-obscured galaxies (BluDOGs) at lower redshifts. 

Nevertheless, type-I AGNs are not the only plausible scenario that can explain the observational features of LRDs. Strong outflows in compact regions, likely driven by supernovae or star formation, can also generate broad H$\alpha$ emission lines. This possibility has been discussed in, e.g., \citet{matthee24}. \textcolor{black}{\citet{baggen24} proposed that the broad lines of LRDs can be explained by the high central densities of these compact objects. Nevertheless,} \textcolor{black}{most spectroscopically-observed LRDs show narrow unresolved forbidden lines (like {\oiii}), which make it hard to explain the broad H$\alpha$ lines purely with galactic-scale outflows \citep[e.g.,][]{maiolino24}.}

\textcolor{black}{All these uncertainties prevent}  us from further understanding the properties of LRDs and their implications on galaxy evolution models. 
Confirming the nature of LRDs requires multi-wavelength observations that can unambiguously reveal signs of AGN activity. Some efforts have been carried out using the Mid-Infrared Instrument (MIRI) on {\em JWST}, aiming at detecting the hot dust emission heated by the AGN. 
%% ACE: say here, why we need MIRI, i.e. to look for the hot dust in the torus around the AGN. 
For example, \citet{pg24} analyzed the NIRCam and MIRI observations of 31 photometrically-selected LRDs and performed SED fitting to test the AGN hypothesis. The authors found that the mid-infrared emissions of the LRDs are inconsistent with an AGN dust torus, potentially indicating that the SEDs of LRDs might not be AGN-dominated. 

%To determine
X-ray emission is one of the most reliable indicators of AGN activity. %However, LRDs seems to be faint in X-ray.
Several studies have investigated the X-ray emission of LRDs in the {\em Chandra} deep fields \citep[e.g.,][]{matthee24, lyu23}, who reported non-detections for all the LRDs in their sample. \citet{wang24} reported a luminous LRD at $z=3.1$ with an X-ray flux upper limit that is $\sim100$ times weaker than the expectation from its optical luminosity.
\citet{inayoshi24} indicate that current X-ray observations are not deep enough to detect the X-ray emission for most individual LRDs.
\textcolor{black}{Meanwhile, \citet{kocevski24} recently discovered two bright photometrically-selected LRDs detected in {\em Chandra} deep fields, which have $L_\text{X}(2-10\text{ keV})\sim10^{44}\text{ erg s}^{-1}$ and $N_\text{H}\sim10^{23}\text{cm}^{-2}$. The diverse results demonstrate that our knowledge about the X-ray properties of LRDs are still highly limited.}

% In this case, stacking analysis offer a viable way to detect possible X-ray emissions of the whole LRD population and test the AGN hypothesis. 
%These studies implied upper limits of the X-ray luminosities to be ...
%To put stronger constraints on the X-ray properties of LRDs, we carry out stacking analysis of 

Since most \textcolor{black}{spectroscopically-confirmed} LRDs do not have X-ray detections, a stacking analysis offers a viable way to investigate the X-ray properties of these objects.
In this letter, we aim to detect or put more stringent constraints on the X-ray fluxes of LRDs by stacking archival {\em Chandra} observations.
We describe our sample and the stacking procedure in Section \ref{sec:analysis}. The results are presented in Section \ref{sec:results}. We discuss the implication of our results in Section \ref{sec:discussion} and conclude in Section \ref{sec:conclusions}. We adopt a flat $\Lambda$CDM cosmology with $H_0=70\text{ km s}^{-1}\text{ Mpc}^{-1}$ and $\Omega_M=0.3$.

%\section{Sample} \label{sec:sample}

\section{Stacking the Chandra Observations of Little Red Dots} \label{sec:analysis}

\subsection{Sample Selection} \label{sec:sample}

\textcolor{black}{The aim of this study is to investigate the AGN hypothesis of LRDs. Therefore, our sample consists of spectroscopically-confirmed LRDs that show broad H$\alpha$ emission lines. We notice that some previous studies reported photometrically-selected LRDs \citep[e.g.,][]{williams23,kokorev24} and discussed their properties; however, these objects lack spectroscopic confirmation and we do not include these objects in our analysis. }

\textcolor{black}{Specifically,
we construct our sample by selecting broad H$\alpha$ emitters (with FWHM$_{\text{H}\alpha}>1000\text{ km s}^{-1}$) discovered by {\em JWST} NIRCam grism or NIRSpec observations by the time of writing this paper.
%Specifically, we select objects that fall in {\em Chandra} deep fields and exhibit broad (FWHM$>1000\text{ km s}^{-1}$) H$\alpha$ emission lines in their {\em JWST} NIRCam or NIRSpec spectra.
We further require that the targets should fall in one of the following {\em Chandra} deep fields: COSMOS \citep{cosmos}, CDF-N \citep{cdfn}, CDF-S \citep{cdfs}, AEGIS \citep{aegis}, and X-UDS \citep{xuds}. This choice is made because the X-ray stacking software we use, {{CSTACK}} \citep{cstack}, is currently only implemented to work on ACIS-I observations in {\em Chandra} deep fields. 
Objects in lensing cluster fields \citep[e.g., those in ][]{greene23} are not covered by the database of CSTACK and are thus not included in our sample. Since these objects might suffer contamination from the X-ray emission of the foreground galaxy cluster, excluding these objects also avoids possible systematic uncertainties due to imperfect background subtraction.}

\textcolor{black}{During the stacking process (Section \ref{sec:stack}), we further exclude five objects with close companions or uneven background by visual inspection, for which the background subtraction might be inaccurate. The final sample of this work consists of 34 objects from \citet{harikane23}, \citet{matthee24}, \citet{maiolino23}, and \citet{kocevski24}. All information about these objects is summarized in Table \ref{tbl:sample}.}

%We also note that the stacking software we use, {{CSTACK}} \citep{cstack}, is currently only implemented to work on ACIS-I observations in {\em Chandra} deep fields,
%and its dataset does not cover lensing cluster fields. 
%The parent sample consists of 
%Our parent sample consists of
%59 objects reported in \cite{harikane23}, \cite{maiolino23}, \cite{matthee24}, and \citet{kocevski24}. 
%We note that some objects in the parent sample were also reported by other studies \citep[e.g.,][]{kocevski23}. }%We also note that \citet{kocevski24} recently reported 17 new LRDs with spectroscopically-confirmed broad H$\alpha$ lines; however, we do not include these objects as their H$\alpha$ luminosities were not available, which is a key part in our analysis (see Section \ref{sec:results}).}
%that show broad (FWHM$>1000\text{ km s}^{-1}$) H$\alpha$ emission lines in their {\em JWST} NIRCam or NIRSpec spectra. 

%\textcolor{black}{
%We then select objects in the {\em Chandra} deep fields: COSMOS \citep{cosmos}, CDF-N \citep{cdfn}, CDF-S \citep{cdfs}, AEGIS \citep{aegis}, and X-UDS \citep{xuds}.
%This choice is made because the stacking software we use, {{CSTACK}} \citep{cstack}, is currently only implemented to work on ACIS-I observations in {\em Chandra} deep fields.  }

\textcolor{black}{Note that we did not apply any additional color cuts when putting together this sample of LRDs. In contrast, previous studies have employed various color cuts in constructing their LRD samples. In a recent study, \citet{kocevski24} provided a comprehensive discussion on LRD color selection and compiled a photometrically-selected sample with homogeneous color cuts across multiple {\em JWST} surveys. Their sample selection, described in Section 3.1 of \citet{kocevski24}, requires objects to have blue SED slopes in the rest-frame UV and red SED slopes in the rest-frame optical. We find that 18 of the LRDs from \citet{kocevski24} are included in our sample. These 18 objects form a sample of spectroscopically-confirmed LRDs with homogeneous color properties
and enable us to assess the impact of sample selection on our results. We refer to these objects as the ``color-selected subset" throughout this study.} %When presenting the results, we will consider both the entire sample (i.e., xxx objects in Table 1) and the K24 subset.}

\begin{deluxetable*}{c|ccccccc}
\label{tbl:sample}
\tablecaption{The LRD sample for stacking}
\tablewidth{0pt}
\tablehead{\colhead{Name} & \colhead{RA} & \colhead{Dec} &  \colhead{Redshift} & \colhead{$\log L_{\text{H}\alpha}$} & \colhead{$\text{ FWHM}_{\text{H}\alpha}$} &  \colhead{$\log M_\text{BH}$\tablenotemark{1}} & \colhead{Reference\tablenotemark{2}}\\
\colhead{} & \colhead{(deg)} & \colhead{(deg)} & \colhead{} & \colhead{(erg s$^{-1}$)} & \colhead{$\text{km s}^{-1}$} & \colhead{$(M_\odot)$} & \colhead{}}
\startdata
\hline
CEERS-00397 & 214.83621 & 52.88269 & 6.000 & $42.41^{+0.12}_{-0.07}$ & $1739^{+359}_{-317}$ & $7.00^{+0.26}_{-0.30}$ & \citet{harikane23}\\
CEERS-00672\tablenotemark{$\dagger$} & 214.88967 & 52.83297 & 5.666 & $43.26^{+0.05}_{-0.05}$ & $2208^{+277}_{-241}$ & $7.70^{+0.13}_{-0.13}$ & \citet{harikane23}\\
CEERS-00717 & 215.08142 & 52.97219 & 6.936 & $42.08^{+0.10}_{-0.08}$ & $6279^{+805}_{-881}$ & $7.99^{+0.16}_{-0.17}$ & \citet{harikane23}\\
CEERS-00746\tablenotemark{$\dagger$} & 214.80912 & 52.86847 & 5.624 & $43.83^{+0.02}_{-0.03}$ & $1660^{+157}_{-162}$ & $7.76^{+0.10}_{-0.11}$ & \citet{harikane23}\\
CEERS-01236 & 215.14529 & 52.96728 & 4.484 & $41.68^{+0.09}_{-0.08}$ & $3521^{+649}_{-485}$ & $7.26^{+0.19}_{-0.18}$ & \citet{harikane23}\\
CEERS-01244 & 215.24067 & 53.03606 & 4.478 & $42.89^{+0.01}_{-0.01}$ & $2228^{+75}_{-52}$ & $7.51^{+0.04}_{-0.03}$ & \citet{harikane23}\\
CEERS-01665 & 215.17821 & 53.05936 & 4.483 & $42.83^{+0.05}_{-0.04}$ & $1794^{+282}_{-171}$ & $7.28^{+0.15}_{-0.13}$ & \citet{harikane23}\\
CEERS-02782 & 214.82346 & 52.83028 & 5.241 & $42.88^{+0.04}_{-0.04}$ & $2534^{+260}_{-266}$ & $7.62^{+0.11}_{-0.12}$ & \citet{harikane23}\\
GOODS-N-13733 & 189.05708 & 62.26894 & 5.236 & $42.38^{+0.03}_{-0.04}$ & $2208^{+200}_{-200}$ & $7.49^{+0.10}_{-0.10}$ & \citet{matthee24}\\
GOODS-N-16813 & 189.17929 & 62.29253 & 5.355 & $42.65^{+0.05}_{-0.05}$ & $2033^{+219}_{-219}$ & $7.55^{+0.12}_{-0.12}$ & \citet{matthee24}\\
GOODS-N-4014 & 189.30013 & 62.21204 & 5.228 & $42.66^{+0.02}_{-0.02}$ & $2103^{+159}_{-159}$ & $7.58^{+0.08}_{-0.08}$ & \citet{matthee24}\\
GOODS-S-13971 & 53.13858 & -27.79025 & 5.481 & $42.40^{+0.08}_{-0.10}$ & $2192^{+479}_{-479}$ & $7.49^{+0.25}_{-0.25}$ & \citet{matthee24}\\
M23-10013704-2\tablenotemark{$\dagger$} & 53.12654 & -27.81809 & 5.919 & $41.99^{+0.03}_{-0.04}$ & $2416^{+179}_{-156}$ & $7.50^{+0.31}_{-0.31}$ & \citet{maiolino23}\\
M23-20621 & 189.12252 & 62.29285 & 4.681 & $42.00^{+0.03}_{-0.03}$ & $1638^{+148}_{-150}$ & $7.30^{+0.31}_{-0.31}$ & \citet{maiolino23}\\
M23-3608 & 189.11794 & 62.23552 & 5.269 & $41.46^{+0.10}_{-0.09}$ & $1373^{+361}_{-198}$ & $6.82^{+0.38}_{-0.33}$ & \citet{maiolino23}\\
M23-53757-2 & 189.26978 & 62.19421 & 4.448 & $42.03^{+0.04}_{-0.04}$ & $2416^{+179}_{-157}$ & $7.69^{+0.32}_{-0.31}$ & \citet{maiolino23}\\
M23-73488-2 & 189.19740 & 62.17723 & 4.133 & $42.52^{+0.01}_{-0.01}$ & $2160^{+45}_{-46}$ & $7.71^{+0.30}_{-0.30}$ & \citet{maiolino23}\\
M23-77652 & 189.29323 & 62.19900 & 5.229 & $42.09^{+0.02}_{-0.04}$ & $1070^{+219}_{-180}$ & $6.86^{+0.35}_{-0.34}$ & \citet{maiolino23}\\
M23-8083 & 53.13284 & -27.80186 & 4.648 & $41.92^{+0.02}_{-0.02}$ & $1648^{+127}_{-130}$ & $7.25^{+0.31}_{-0.31}$ & \citet{maiolino23}\\
CEERS-5760\tablenotemark{$\dagger$} & 214.97244 & 52.96220 & 5.079 & $41.81^{+0.15}_{-0.15}$ & $1800^{+300}_{-300}$ & $6.72^{+0.17}_{-0.17}$ & \citet{kocevski24}\\
CEERS-6126\tablenotemark{$\dagger$} & 214.92338 & 52.92559 & 5.288 & $42.76^{+0.03}_{-0.03}$ & $1670^{+60}_{-60}$ & $7.18^{+0.03}_{-0.03}$ & \citet{kocevski24}\\
CEERS-7902\tablenotemark{$\dagger$} & 214.88680 & 52.85538 & 6.986 & $43.20^{+0.07}_{-0.07}$ & $4180^{+220}_{-220}$ & $8.24^{+0.06}_{-0.06}$ & \citet{kocevski24}\\
CEERS-10444\tablenotemark{$\dagger$} & 214.79537 & 52.78885 & 6.687 & $43.38^{+0.06}_{-0.06}$ & $5420^{+370}_{-370}$ & $8.57^{+0.07}_{-0.07}$ & \citet{kocevski24}\\
CEERS-13135\tablenotemark{$\dagger$} & 214.99098 & 52.91652 & 4.955 & $42.98^{+0.03}_{-0.03}$ & $1850^{+50}_{-50}$ & $7.39^{+0.03}_{-0.03}$ & \citet{kocevski24}\\
CEERS-13318\tablenotemark{$\dagger$} & 214.98304 & 52.95601 & 5.280 & $43.77^{+0.05}_{-0.05}$ & $3300^{+60}_{-60}$ & $8.34^{+0.03}_{-0.03}$ & \citet{kocevski24}\\
CEERS-14949\tablenotemark{$\dagger$} & 215.13706 & 52.98856 & 5.684 & $42.50^{+0.08}_{-0.08}$ & $1520^{+115}_{-115}$ & $6.95^{+0.08}_{-0.08}$ & \citet{kocevski24}\\
CEERS-20496\tablenotemark{$\dagger$} & 215.07826 & 52.94850 & 6.786 & $41.71^{+0.23}_{-0.23}$ & $1410^{+200}_{-200}$ & $6.45^{+0.18}_{-0.18}$ & \citet{kocevski24}\\
CEERS-20777\tablenotemark{$\dagger$} & 214.89225 & 52.87741 & 5.286 & $42.13^{+0.03}_{-0.03}$ & $1690^{+70}_{-70}$ & $6.84^{+0.04}_{-0.04}$ & \citet{kocevski24}\\
PRIMER-UDS-29881\tablenotemark{$\dagger$} & 34.31313 & -5.22677 & 6.170 & $41.89^{+0.12}_{-0.12}$ & $2280^{+220}_{-220}$ & $6.98^{+0.11}_{-0.11}$ & \citet{kocevski24}\\
PRIMER-UDS-31092\tablenotemark{$\dagger$} & 34.26458 & -5.23254 & 5.675 & $42.20^{+0.05}_{-0.05}$ & $2160^{+130}_{-130}$ & $7.10^{+0.06}_{-0.06}$ & \citet{kocevski24}\\
PRIMER-UDS-32438\tablenotemark{$\dagger$} & 34.24181 & -5.23940 & 3.500 & $41.80^{+0.04}_{-0.04}$ & $2420^{+120}_{-120}$ & $6.98^{+0.05}_{-0.05}$ & \citet{kocevski24}\\
PRIMER-UDS-33823\tablenotemark{$\dagger$} & 34.24419 & -5.24583 & 3.103 & $43.76^{+0.02}_{-0.02}$ & $2700^{+130}_{-130}$ & $8.16^{+0.04}_{-0.04}$ & \citet{kocevski24}\\
PRIMER-UDS-116251\tablenotemark{$\dagger$} & 34.26054 & -5.20912 & 5.365 & $41.55^{+0.21}_{-0.21}$ & $1540^{+280}_{-280}$ & $6.44^{+0.20}_{-0.20}$ & \citet{kocevski24}\\
PRIMER-UDS-119639\tablenotemark{$\dagger$} & 34.31214 & -5.20255 & 6.518 & $42.15^{+0.09}_{-0.09}$ & $2200^{+400}_{-400}$ & $7.09^{+0.17}_{-0.17}$ & \citet{kocevski24}\\
\hline
\enddata
%\tablenotetext{1}{The absolute magnitude at rest-frame $1450${\AA}.}
\tablenotetext{1}{The SMBH masses are calculated using the H$\alpha$ broad emission line. The errors in the table only contain statistical errors; the systematic errors of the black hole mass is about 0.3 dex.}
\tablenotetext{2}{The references from which the objects are gathered.}
\tablenotetext{\dagger}{Objects in the photometrically-selected LRD sample in \citet{kocevski24} (i.e., the color selected sample defined in Section \ref{sec:sample}.)}
%\tablenotetext{*}{\citet{kocevski24} did not provide the H$\alpha$ luminosities of their objects. The H$\alpha$ luminosities quoted here are computed using the SMBH masses and H$\alpha$ line widths provided by \citet{kocevski24}, according to their Equation 5.}
%\tablenotetext{3}{References for the quasar properties.}
\tablecomments{\citet{kocevski24} did not provide the H$\alpha$ luminosities of their objects. The H$\alpha$ luminosities quoted here for objects fro \citet{kocevski24} are computed using the SMBH masses and H$\alpha$ line widths provided by \citet{kocevski24}, according to their Equation 5.}
\end{deluxetable*}
%\end{longrotatetable}

\subsection{Stacking Analysis} \label{sec:stack}

%We use two softwares to stack the archival ACIS-I observations.
%There are two online tools to stack archival ACIS-I observations: \texttt{CSTACK} \citep{cstack} and \texttt{stackfast} (Annana et al. in prep). In this work, we use both tools to perform stacking analysis, which allow us to cross-check the results.

%\subsection{\texttt{CSTACK}}

We use {CSTACK} \citep{cstack}\footnote{http://lambic.astrosen.unam.mx/cstack/} to analyze the archival {\em{Chandra}} observations. {CSTACK} is an online tool %developed by Takamitsu Miyaji 
for stacking the archival ACIS-I observations in several {\em Chandra} deep fields. 
%% add citation?
\textcolor{black}{Our targets are covered by the CDFN (2Ms), CDFS (7Ms), AEGIS-XD, and X-UDS datasets in CSTACK.}
%To compute the stacked flux for a sample of objects, 
{CSTACK} takes the positions of objects as inputs and sums the photons within certain apertures at these positions. \textcolor{black}{In this work, we use circular apertures that correspond to an enclosed counts fraction (ECF) of 0.9, and
 only include observations where the target has an off-axis angle $(\theta_\text{off})$ smaller than $8'$.}  
%{More than 80\% of the objects have aperture radii smaller than $3''$. }%, and the maximum aperture radii is $5\farcs8$.}
%{CSTACK} also estimates and subtracts the background level. 
\textcolor{black}{The background levels of observations are estimated using the square regions centered on the source positions with sizes of $30''\times30''$, excluding the inner circular regions with a radius of $7''$.}
The background level is then subtracted from the source flux.
The output of {CSTACK} includes the stacked photon count rates and their errors in the soft $(0.5-2\text{ keV})$ and hard $(2-8\text{ keV})$ X-ray bands as well as the stacked images \textcolor{black}{in the two bands}.% The count rates can be converted to physical fluxes 

%Given that the LRDs in our sample have different redshifts and luminosities, 
%We perform stacking analysis for both individual objects and for the whole sample. 
\textcolor{black}{
To exclude objects not suitable for stacking analysis (e.g., those with bright companions) and to enable a flexible analysis of their physical properties, we run {{CSTACK}} first for individual LRDs instead of the whole sample. We visually inspect the images of individual objects and exclude \textcolor{black}{five objects} with bright companions or uneven background, which can cause inaccurate background estimation and subtraction. \textcolor{black}{For the remaining objects, we obtain the photon counts in source and background regions in individual {\em Chandra} observations from CSTACK}, and use these quantities to derive the posterior probability of the count rate for each source and for the sample stack (Section \ref{sec:prob}).} %We then run {{CSTACK}} again for the whole sample to get the stacked images. 

%\textcolor{black}{To illustrate the result of stacking, in Figure \ref{fig:stackimage}, we present the stacked images of the whole sample produced by CSTACK. The images show no signs of emissions at their centers. Note that Figure 1 itself does not imply a non-detection of the stacking analysis, due to the very different depths and PSF sizes of {\em Chandra} observations. In this work, the stacked count rates are computed by averaging the count rates of individual objects, which we describe in Section \ref{sec:prob}.}

\textcolor{black}{Another factor to consider for stacking X-ray observations is the off-axis angle of objects. The PSF of {\em Chandra} strongly depends on the off-axis angle; according to the {\em Chandra} Proposers’ Observatory Guide\footnote{https://cxc.harvard.edu/proposer/POG/pdf/MPOG.pdf}, the PSF of {\em Chandra} only changes subtly at $\theta_\text{off}<2'$ and quickly becomes fatter and more eccentric at larger radii. Fat and eccentric PSFs can introduce systematic errors in aperture correction and background subtraction. In {{CSTACK}}, this effect is accounted for by performing photometry using apertures corresponding to ECF$=0.9$. \textcolor{black}{By requiring $\theta_\text{off}<8'$, more than 80\% of the observations we analyzed have aperture radii smaller than $3''$.} To further investigate the impact of large off-axis angles, we construct another subset of LRD observations with object off-axis angle $\theta_\text{off}<2'$, which is referred to as ``2-arcmin subset" in the rest of this paper. We will compare the result for this subset and the whole sample in Section \ref{sec:results}.}

%\begin{figure}
%    \centering
%    \includegraphics[width=1\linewidth]{stackedimage_cstack.pdf}
%    \caption{The stacked images (size $30''\times30''$) of LRDs in the soft (\textit{left}) and hard (\textit{right}) X-ray band produced by {CSTACK}. \textcolor{black}{The total exposure time is 26.6 Ms.} The yellow circles (with radii of $2''$) mark the source position. \textcolor{black}{Note that this figure illustrates the output stacked images from CSTACK, though we do not use the stacked images to estimate the upper limit of the stacked fluxes (see Section \ref{sec:prob} for details).}}
%    \label{fig:stackimage}
%\end{figure}
%% ACE: add colorbar to this figure and mark central position by a cross (or a circle with a typical or something like that

%% ACE: maybe add new subsection here
\subsection{X-ray \textcolor{black}{Count Rate} Limits} \label{sec:prob}

We then evaluate the upper limit of the photon rates for individual LRDs and for the whole sample. Since our targets only have a few (or zero) photons in their source apertures, 
%% ACE: wait, so for some of them you detect photons? A few photons is a lot, right? Some of the bright type I quasars have only a few photons... and why only majority? For some you detect a lot of photons? 
we need to use Poisson statistics instead of Gaussian distributions to describe the photon rates. We note that CSTACK estimates the error of the stacked flux by bootstrapping the sample, which does not work for individual objects. 
\textcolor{black}{To derive the posterior distribution of photon rates for individual objects, we follow the algorithm used by the {\texttt{aprates}}\footnote{The detailed description of the algorithm can be found at {https://cxc.harvard.edu/csc1/memos/files/Kashyap\_xraysrc.pdf}} task in CIAO \citep{ciao}. We briefly describe this algorithm below.}

For the $i$th image to be stacked, CSTACK computes the photon rate of a source as follows:
\begin{equation} \label{eq:rs}
    r_{s,i} = \left[\frac{N_i}{t_{i}} - \frac{N_{b,i}A_{s,i}}{t_{b,i}A_{b,i}}\right]/\text{ECF}
\end{equation}
where $N_i$ $(N_{b,i})$ is the number of photons within the source aperture (within the background region),
$t_{i}$ $(t_{b,i})$ is the exposure time on the source (on the background region), and $A_{s,i}$ $(A_{b,i})$ is the area of the source aperture (background region).
%The factor 0.9 accounts for the ECF of the apertures.
\textcolor{black}{The stacked photon rate, $r_s$, can then be computed as
\begin{equation} \label{eq:rsavg}
    r_s = \frac{\sum_i t_ir_{s,i}}{\sum_i t_i} = \frac{\sum_i N_i}{\text{ECF}\sum_i t_i}-\frac{1}{\text{ECF}\sum_i t_i}\sum_i \frac{N_{b,i}t_iA_{s,i}}{t_{b,i}A_{b,i}}
\end{equation}}

CSTACK delivers the observed values of $N_i$ and $N_{b,i}$, which we denote as $N_i^\text{obs}$ and $N_{b,i}^\text{obs}$. These photon counts follow the Poisson distribution. In the following text, we use $N_i$ and $N_{b,i}$ to denote the expectation of the Poisson distributions. According to Bayes' Theorem, the probability for the observation to have an expected photon count of $N_i$ in the source aperture is
\begin{equation} \label{eq:ns}
    P(N_i|N_i^\text{obs}) = \frac{P(N_i^\text{obs}|N_i) P(N_i)}{P(N_i^\text{obs})}
     = \frac{N_i^{N_i^\text{obs}}e^{-N_i}}{N_i^\text{obs}!}
\end{equation}
For the second equal sign, we assume an \textcolor{black}{uninformative} flat prior for $N_i$ and normalize the probabilistic distribution.
Similarly, for $N_{b,i}$ we have 
\begin{equation} \label{eq:nb}
    P(N_{b,i}|N_{b,i}^\text{obs}) = \frac{N_{b,i}^{N_{b,i}^\text{obs}}e^{-N_{b,i}}}{N_{b,i}^\text{obs}!}
\end{equation}
\textcolor{black}{
The probabilistic distribution of $r_s$ can be determined by combining Equation \ref{eq:rs}, \ref{eq:rsavg}, \ref{eq:ns}, and \ref{eq:nb}.
We also compute the full band ($0.5-8$ keV) by combining the soft and hard band count rate distributions.} 

%Using this method, we compute the probabilistic distribution of the true photon rates for individual objects and for the whole stacked sample. %The values and errors reported in Table \ref{tbl:stack} correspond to the 16, 50, and 84 percentile of the distribution. The upper limits we used throughout this Letter correspond to a cumulative probability of 99.87\%, equivalent to a $3\sigma$ upper limit in Normal distributions.

Table \ref{tbl:stack} lists the evaluated \textcolor{black}{count rates for individual objects and for the whole sample. None of the LRDs are detected individually in either the soft or hard band with more than $2.6\sigma$ significance.} \textcolor{black}{After stacking all objects in our sample, we get tentative detections in the soft and the hard band with $2.9\sigma$ and $3.2\sigma$ significance. Combining the two bands gives a $4.1\sigma$ detection in the full band for the whole sample stack. This result indicates that AGN activity likely exists in the broad H$\alpha$ emitters in our sample. The color-selected subset and the 2-arcmin subset do not show significant detections ($>3\sigma$) in any band, possibly due to shorter stacked exposure time for the subsets.} 

In the rest of this paper, we define upper limits as the values corresponding to a cumulative probability of 99.87\%. This definition is equivalent to a $3\sigma$ upper limit in a Gaussian distribution.

\begin{deluxetable*}{c|ccccccc}
\label{tbl:stack}
%\centerwidetable
\tablecaption{Photon rates for individual objects and the stacked whole sample}
\tablewidth{0pt}
\tablehead{\colhead{Name} & \colhead{Field} & \colhead{$N_H^\text{MW}$\tablenotemark{1}} & \colhead{$t_{\text{exp}}$\tablenotemark{2}} & \colhead{$\overline{\theta_\text{off}}$\tablenotemark{3}} & \colhead{$\text{Rate}_\text{soft}$} & \colhead{$\text{Rate}_\text{hard}$} & \colhead{$\text{Rate}_\text{full}$} \\
\colhead{}  & \colhead{} & \colhead{$(10^{20}\text{cm}^{-2})$} & \colhead{(Ms)} & \colhead{(arcmin)} & \colhead{$(10^{-6}\text{cts s}^{-1})$} & \colhead{$(10^{-6}\text{cts s}^{-1})$} & \colhead{$(10^{-6}\text{cts s}^{-1})$}}
\startdata
\hline
CEERS-00397 & AEGIS & 0.94 & 0.69 & 4.1 &$1.31^{+4.39}_{-3.37}$ & $-2.07^{+6.81}_{-5.81}$ &$-0.26^{+7.98}_{-6.96}$ \\
CEERS-00672 & AEGIS & 0.95 & 0.71 & 2.9 &$-1.40^{+1.77}_{-0.82}$ & $1.71^{+5.19}_{-4.17}$ &$0.75^{+5.44}_{-4.42}$ \\
CEERS-00717 & AEGIS & 0.95 & 0.67 & 2.4 &$-0.59^{+1.89}_{-0.87}$ & $1.92^{+4.56}_{-3.47}$ &$1.82^{+4.89}_{-3.80}$ \\
CEERS-00746 & AEGIS & 0.92 & 0.69 & 3.4 &$-0.55^{+2.60}_{-1.58}$ & $5.26^{+6.26}_{-5.22}$ &$5.21^{+6.69}_{-5.66}$ \\
CEERS-01236 & AEGIS & 0.94 & 0.70 & 3.8 &$1.46^{+3.55}_{-2.53}$ & $1.81^{+6.03}_{-5.01}$ &$3.77^{+6.88}_{-5.86}$ \\
CEERS-01244 & AEGIS & 0.93 & 0.67 & 6.2 &$-6.61^{+6.27}_{-5.38}$ & $8.79^{+13.23}_{-12.32}$ &$2.61^{+14.53}_{-13.63}$ \\
CEERS-01665 & AEGIS & 0.93 & 0.70 & 4.7 &$11.04^{+6.28}_{-5.30}$ & $-9.80^{+8.12}_{-7.29}$ &$1.70^{+10.13}_{-9.24}$ \\
CEERS-02782 & AEGIS & 0.93 & 0.72 & 1.5 &$1.79^{+2.49}_{-1.50}$ & $-1.06^{+3.05}_{-2.05}$ &$1.21^{+3.86}_{-2.85}$ \\
GOODS-N-13733 & CDF-N & 0.96 & 1.64 & 4.2 &$0.13^{+2.90}_{-2.49}$ & $4.87^{+5.16}_{-4.75}$ &$5.20^{+5.86}_{-5.45}$ \\
GOODS-N-16813 & CDF-N & 0.98 & 1.20 & 4.0 &$-1.77^{+2.10}_{-1.53}$ & $8.30^{+5.40}_{-4.80}$ &$6.80^{+5.74}_{-5.14}$ \\
GOODS-N-4014 & CDF-N & 0.98 & 1.59 & 3.5 &$1.36^{+2.17}_{-1.72}$ & $0.47^{+3.52}_{-3.08}$ &$2.05^{+4.07}_{-3.63}$ \\
GOODS-S-13971 & CDF-S & 0.69 & 6.74 & 1.5 &$0.35^{+0.44}_{-0.34}$ & $-0.02^{+0.98}_{-0.85}$ &$0.38^{+1.06}_{-0.94}$ \\
M23-10013704-2 & CDF-S & 0.69 & 5.72 & 0.9 &$-0.04^{+0.38}_{-0.25}$ & $-0.20^{+0.99}_{-0.83}$ &$-0.18^{+1.05}_{-0.90}$ \\
M23-20621 & CDF-N & 0.98 & 1.48 & 4.0 &$-3.08^{+1.73}_{-1.29}$ & $3.41^{+4.83}_{-4.36}$ &$0.55^{+5.10}_{-4.63}$ \\
M23-3608 & CDF-N & 0.98 & 1.05 & 2.6 &$2.63^{+2.62}_{-1.94}$ & $2.16^{+4.02}_{-3.33}$ &$5.13^{+4.71}_{-4.02}$ \\
M23-53757-2 & CDF-N & 0.97 & 1.48 & 3.4 &$5.17^{+2.82}_{-2.33}$ & $8.81^{+4.52}_{-4.04}$ &$14.22^{+5.26}_{-4.77}$ \\
M23-73488-2 & CDF-N & 0.97 & 1.50 & 3.6 &$1.62^{+2.00}_{-1.52}$ & $2.37^{+3.44}_{-2.97}$ &$4.23^{+3.92}_{-3.44}$ \\
M23-77652 & CDF-N & 0.97 & 1.68 & 3.5 &$0.08^{+1.94}_{-1.52}$ & $-0.67^{+3.50}_{-3.09}$ &$-0.39^{+3.95}_{-3.53}$ \\
M23-8083 & CDF-S & 0.69 & 5.70 & 0.8 &$-0.07^{+0.38}_{-0.25}$ & $-0.18^{+0.99}_{-0.84}$ &$-0.19^{+1.05}_{-0.90}$ \\
CEERS-5760 & AEGIS & 0.94 & 0.68 & 4.6 &$2.94^{+4.57}_{-3.62}$ & $8.36^{+9.53}_{-8.70}$ &$11.74^{+10.48}_{-9.62}$ \\
CEERS-6126 & AEGIS & 0.94 & 0.63 & 6.4 &$-8.78^{+5.90}_{-5.05}$ & $24.74^{+14.78}_{-13.85}$ &$16.37^{+15.83}_{-14.91}$ \\
CEERS-7902 & AEGIS & 0.94 & 0.58 & 3.6 &$1.33^{+3.76}_{-2.53}$ & $4.80^{+7.29}_{-6.05}$ &$6.74^{+8.08}_{-6.83}$ \\
CEERS-10444 & AEGIS & 0.93 & 0.58 & 1.9 &$3.72^{+3.71}_{-2.48}$ & $1.21^{+4.79}_{-3.56}$ &$5.55^{+5.93}_{-4.69}$ \\
CEERS-13135 & AEGIS & 0.95 & 0.63 & 5.9 &$-3.93^{+5.61}_{-4.60}$ & $8.15^{+11.41}_{-10.37}$ &$4.71^{+12.59}_{-11.56}$ \\
CEERS-13318 & AEGIS & 0.95 & 0.61 & 4.6 &$5.33^{+5.69}_{-4.58}$ & $-3.05^{+9.48}_{-8.45}$ &$2.81^{+10.91}_{-9.86}$ \\
CEERS-14949 & AEGIS & 0.94 & 0.66 & 2.8 &$-0.50^{+1.93}_{-0.88}$ & $6.45^{+5.55}_{-4.44}$ &$6.45^{+5.83}_{-4.72}$ \\
CEERS-20496 & AEGIS & 0.95 & 0.71 & 3.7 &$0.34^{+3.06}_{-2.08}$ & $4.84^{+6.62}_{-5.63}$ &$5.66^{+7.20}_{-6.21}$ \\
CEERS-20777 & AEGIS & 0.94 & 0.59 & 4.5 &$7.10^{+6.20}_{-5.00}$ & $-5.75^{+8.63}_{-7.49}$ &$1.93^{+10.46}_{-9.30}$ \\
PRIMER-UDS-29881 & X-UDS & 2.01 & 0.33 & 5.1 &$-7.38^{+6.89}_{-5.18}$ & $-0.01^{+14.95}_{-13.17}$ &$-6.56^{+16.33}_{-14.56}$ \\
PRIMER-UDS-31092 & X-UDS & 1.98 & 0.29 & 3.9 &$22.12^{+13.72}_{-11.57}$ & $7.16^{+16.78}_{-15.27}$ &$30.20^{+21.44}_{-19.69}$ \\
PRIMER-UDS-32438 & X-UDS & 1.98 & 0.23 & 3.7 &$0.88^{+7.81}_{-4.82}$ & $4.41^{+12.86}_{-9.94}$ &$6.71^{+14.80}_{-11.85}$ \\
PRIMER-UDS-33823 & X-UDS & 1.98 & 0.26 & 3.7 &$3.00^{+8.32}_{-5.69}$ & $5.82^{+12.91}_{-10.39}$ &$10.09^{+15.10}_{-12.54}$ \\
PRIMER-UDS-116251 & X-UDS & 2.04 & 0.32 & 4.6 &$14.95^{+10.31}_{-8.22}$ & $-1.64^{+13.33}_{-11.80}$ &$14.23^{+16.62}_{-14.89}$ \\
PRIMER-UDS-119639 & X-UDS & 2.01 & 0.42 & 5.5 &$9.05^{+9.29}_{-7.72}$ & $5.58^{+14.48}_{-13.17}$ &$15.36^{+17.00}_{-15.61}$ \\\hline\hline
Whole Sample & - & - & 42.86 & 2.7 & $1.03_{-0.37}^{+0.38}$ & $2.37_{-0.76}^{+0.78}$ & $3.21
_{-0.80}^{+0.82}$ \\
Color-selected subset & - & - & 14.66 & 2.9 & $1.48_{-0.72}^{+0.77}$ & $3.59_{-1.54}^{+1.58}$ & $4.79_{-1.60}^{+1.64}$ \\
2-arcmin subset & - & - & 22.30 & 1.1 & $0.67_{-0.23}^{+0.27}$ & $0.37_{-0.49}^{+0.52}$ & $1.00_{-0.54}^{+0.58}$ 
\enddata
\tablenotetext{1}{The column density of the Milky Way absorption.}
\tablenotetext{2}{The total effective exposure time reported by CSTACK.}
\tablenotetext{3}{The mean off-axis angle of the object, weighted by the exposure time of observations.}
\tablecomments{For count rates, the values are the 16th, 50th, and 84th percentiles of the distributions.
The ``soft", ``hard", and ``full" bands stand for 0.5-2 keV, 2-8 keV, and 0.5-8 keV, respectively.}
%\tablecomments{Information about the individual observations used in stacking is available online at (addr.)}
\end{deluxetable*}

\subsection{Estimating Physical Fluxes and Luminosities}

%As the next step, %we convert photon rates to physical fluxes.
 We convert photon rates (in counts$\text{ s}^{-1}$) to fluxes (in $\text{erg s}^{-1}\text{cm}^{-2}$) of the LRDs using the {\em Chandra} Proposal Planning Toolkit\footnote{https://cxc.harvard.edu/toolkit/pimms.jsp}.
\textcolor{black}{Note that the sensitivity of {\em Chandra} degrades with time. To account for this effect, CSTACK rescales the exposure maps of archival {\em Chandra} observations to match the sensitivity of certain {\em Chandra} Cycles. Following the manual of {{CSTACK}}, we use the conversion factors of Cycle 10 for CDF-S, Cycle 3 for CDF-N, Cycle 9 for AEGIS, and Cycle 16 for X-UDS.}
We use a power law with photon index $\Gamma=1.8$ to describe the unabsorbed X-ray spectra of LRDs. \textcolor{black}{To evaluate the impact of AGN photoelectric absorption, we consider three different column densities, i.e., $N_H=10^{21}, 10^{22},\text{ and } 10^{23}\text{ cm}^{-2}$. We note that $N_H=10^{22}\text{ cm}^{-2}$ is usually considered the boundary between Type-I and Type-II AGNs  \citep[e.g.,][]{koss17, ricci17, panagiotou19}.} \textcolor{black}{We use the online HEASoft tool\footnote{https://heasarc.gsfc.nasa.gov/cgi-bin/Tools/w3nh/w3nh.pl} to obtain the column densities of the Milky Way absorption, where we adopt the column density map from \citet{HI4PI}.} 

We also compute the upper limits of the unabsorbed X-ray luminosities at rest-frame 2 to 10 keV ($L_\text{X})$ for the LRDs. \textcolor{black}{For each object, we compute the upper limit of $L_X$ indicated from soft, hard, and total band fluxes, then choose the lowest value as the upper limit of $L_X$. We find that soft band fluxes give the tightest constraint in nearly all cases.} The estimated physical fluxes and luminosity upper limits are listed in Table \ref{tbl:fluxlum}.

All the data and code used in this work, \textcolor{black}{including more detailed information about individual observations used in the stacking analysis and the posterior distribution of the photon count rates,} are available at https://github.com/cosmicdawn-mit/xray\_lrd.

\begin{longrotatetable}
\begin{deluxetable*}{c|cccc|cccc|cccc}
\label{tbl:fluxlum}
%\centerwidetable
\tablecaption{Derived fluxes and luminosity limits of the sample}
\tablewidth{0pt}
\tablehead{\colhead{Name} & \multicolumn{4}{c}{$N_\text{H}=10^{21}\text{cm}^{-2}$} & \multicolumn{4}{c}{$N_\text{H}=10^{22}\text{cm}^{-2}$} & \multicolumn{4}{c}{$N_\text{H}=10^{23}\text{cm}^{-2}$} \\
\colhead{} &
\colhead{$F_\text{soft}$\tablenotemark{1}} & \colhead{$F_\text{hard}$\tablenotemark{2}} & \colhead{$F_\text{full}$\tablenotemark{3}} & \colhead{$L_\text{X}$\tablenotemark{4}} &
\colhead{$F_\text{soft}$} & \colhead{$F_\text{hard}$} & \colhead{$F_\text{full}$} & \colhead{$L_\text{X}$} &
\colhead{$F_\text{soft}$} & \colhead{$F_\text{hard}$} & \colhead{$F_\text{full}$} & \colhead{$L_\text{X}$} }
\startdata
CEERS-00397 & $0.74^{+2.49}_{-1.91}$ & $-3.86^{+12.72}_{-10.84}$ & $-0.49^{+14.89}_{-13.00}$ & $<4.82$ & $0.73^{+2.47}_{-1.89}$ & $-3.86^{+12.72}_{-10.84}$ & $-0.49^{+14.89}_{-13.00}$ & $<4.90$ & $0.69^{+2.32}_{-1.78}$ & $-3.88^{+12.78}_{-10.89}$ & $-0.49^{+14.95}_{-13.05}$ & $<5.72$\\
CEERS-00672 & $-0.79^{+1.00}_{-0.46}$ & $3.18^{+9.68}_{-7.78}$ & $1.41^{+10.16}_{-8.25}$ & $<1.83$ & $-0.79^{+0.99}_{-0.46}$ & $3.18^{+9.68}_{-7.78}$ & $1.41^{+10.16}_{-8.25}$ & $<1.87$ & $-0.74^{+0.94}_{-0.43}$ & $3.20^{+9.72}_{-7.81}$ & $1.41^{+10.20}_{-8.29}$ & $<2.24$\\
CEERS-00717 & $-0.33^{+1.07}_{-0.49}$ & $3.59^{+8.51}_{-6.48}$ & $3.40^{+9.13}_{-7.10}$ & $<3.35$ & $-0.33^{+1.06}_{-0.49}$ & $3.59^{+8.51}_{-6.48}$ & $3.40^{+9.13}_{-7.10}$ & $<3.39$ & $-0.31^{+1.00}_{-0.46}$ & $3.61^{+8.55}_{-6.51}$ & $3.42^{+9.17}_{-7.13}$ & $<3.77$\\
CEERS-00746 & $-0.31^{+1.47}_{-0.89}$ & $9.81^{+11.68}_{-9.75}$ & $9.72^{+12.50}_{-10.57}$ & $<2.56$ & $-0.31^{+1.46}_{-0.89}$ & $9.81^{+11.68}_{-9.75}$ & $9.72^{+12.50}_{-10.57}$ & $<2.61$ & $-0.29^{+1.37}_{-0.84}$ & $9.85^{+11.73}_{-9.79}$ & $9.76^{+12.55}_{-10.61}$ & $<3.13$\\
CEERS-01236 & $0.83^{+2.01}_{-1.43}$ & $3.38^{+11.26}_{-9.35}$ & $7.04^{+12.85}_{-10.93}$ & $<2.22$ & $0.82^{+2.00}_{-1.42}$ & $3.38^{+11.26}_{-9.35}$ & $7.04^{+12.85}_{-10.93}$ & $<2.30$ & $0.77^{+1.88}_{-1.34}$ & $3.39^{+11.31}_{-9.39}$ & $7.07^{+12.90}_{-10.98}$ & $<3.08$\\
CEERS-01244 & $-3.74^{+3.55}_{-3.05}$ & $16.40^{+24.70}_{-23.00}$ & $4.87^{+27.13}_{-25.44}$ & $<2.19$ & $-3.71^{+3.52}_{-3.02}$ & $16.40^{+24.70}_{-23.00}$ & $4.87^{+27.13}_{-25.44}$ & $<2.27$ & $-3.50^{+3.32}_{-2.85}$ & $16.47^{+24.80}_{-23.10}$ & $4.89^{+27.25}_{-25.55}$ & $<3.04$\\
CEERS-01665 & $6.25^{+3.55}_{-3.00}$ & $-18.29^{+15.16}_{-13.61}$ & $3.18^{+18.91}_{-17.25}$ & $<4.68$ & $6.21^{+3.53}_{-2.98}$ & $-18.29^{+15.16}_{-13.61}$ & $3.18^{+18.91}_{-17.25}$ & $<4.84$ & $5.84^{+3.32}_{-2.80}$ & $-18.37^{+15.22}_{-13.67}$ & $3.19^{+18.99}_{-17.33}$ & $<6.18$\\
CEERS-02782 & $1.01^{+1.41}_{-0.85}$ & $-1.99^{+5.70}_{-3.82}$ & $2.26^{+7.20}_{-5.31}$ & $<2.57$ & $1.01^{+1.40}_{-0.84}$ & $-1.99^{+5.70}_{-3.82}$ & $2.26^{+7.20}_{-5.31}$ & $<2.63$ & $0.95^{+1.32}_{-0.79}$ & $-2.00^{+5.72}_{-3.84}$ & $2.27^{+7.23}_{-5.34}$ & $<3.25$\\
GOODS-N-13733 & $0.06^{+1.33}_{-1.14}$ & $10.01^{+10.62}_{-9.77}$ & $10.69^{+12.06}_{-11.21}$ & $<1.64$ & $0.06^{+1.32}_{-1.13}$ & $10.02^{+10.62}_{-9.78}$ & $10.69^{+12.06}_{-11.22}$ & $<1.69$ & $0.06^{+1.27}_{-1.09}$ & $10.05^{+10.66}_{-9.82}$ & $10.73^{+12.11}_{-11.26}$ & $<2.14$\\
GOODS-N-16813 & $-0.81^{+0.96}_{-0.70}$ & $17.06^{+11.10}_{-9.87}$ & $13.99^{+11.80}_{-10.57}$ & $<1.12$ & $-0.81^{+0.96}_{-0.70}$ & $17.07^{+11.11}_{-9.88}$ & $13.99^{+11.81}_{-10.58}$ & $<1.14$ & $-0.78^{+0.92}_{-0.67}$ & $17.13^{+11.15}_{-9.91}$ & $14.05^{+11.85}_{-10.62}$ & $<1.44$\\
GOODS-N-4014 & $0.62^{+0.99}_{-0.79}$ & $0.97^{+7.23}_{-6.33}$ & $4.22^{+8.37}_{-7.46}$ & $<1.50$ & $0.62^{+0.99}_{-0.78}$ & $0.97^{+7.24}_{-6.33}$ & $4.22^{+8.38}_{-7.47}$ & $<1.54$ & $0.60^{+0.95}_{-0.76}$ & $0.97^{+7.27}_{-6.36}$ & $4.24^{+8.41}_{-7.50}$ & $<1.96$\\
GOODS-S-13971 & $0.21^{+0.26}_{-0.20}$ & $-0.05^{+1.90}_{-1.65}$ & $0.75^{+2.06}_{-1.82}$ & $<0.47$ & $0.21^{+0.26}_{-0.20}$ & $-0.05^{+1.90}_{-1.65}$ & $0.75^{+2.06}_{-1.82}$ & $<0.48$ & $0.20^{+0.24}_{-0.19}$ & $-0.05^{+1.91}_{-1.66}$ & $0.75^{+2.07}_{-1.82}$ & $<0.58$\\
M23-10013704-2 & $-0.02^{+0.22}_{-0.15}$ & $-0.38^{+1.92}_{-1.62}$ & $-0.34^{+2.03}_{-1.73}$ & $<0.41$ & $-0.02^{+0.22}_{-0.15}$ & $-0.39^{+1.92}_{-1.62}$ & $-0.34^{+2.03}_{-1.74}$ & $<0.42$ & $-0.02^{+0.21}_{-0.14}$ & $-0.39^{+1.92}_{-1.62}$ & $-0.34^{+2.04}_{-1.74}$ & $<0.50$\\
M23-20621 & $-1.41^{+0.79}_{-0.59}$ & $7.02^{+9.94}_{-8.97}$ & $1.12^{+10.48}_{-9.51}$ & $<0.48$ & $-1.40^{+0.79}_{-0.59}$ & $7.02^{+9.95}_{-8.98}$ & $1.12^{+10.48}_{-9.52}$ & $<0.49$ & $-1.35^{+0.76}_{-0.57}$ & $7.05^{+9.98}_{-9.01}$ & $1.13^{+10.52}_{-9.55}$ & $<0.66$\\
M23-3608 & $1.20^{+1.20}_{-0.89}$ & $4.44^{+8.26}_{-6.85}$ & $10.54^{+9.69}_{-8.27}$ & $<2.09$ & $1.20^{+1.19}_{-0.88}$ & $4.44^{+8.27}_{-6.85}$ & $10.55^{+9.69}_{-8.28}$ & $<2.15$ & $1.16^{+1.15}_{-0.85}$ & $4.46^{+8.30}_{-6.88}$ & $10.59^{+9.73}_{-8.31}$ & $<2.72$\\
M23-53757-2 & $2.37^{+1.29}_{-1.06}$ & $18.12^{+9.30}_{-8.30}$ & $29.25^{+10.81}_{-9.81}$ & $<1.71$ & $2.36^{+1.28}_{-1.06}$ & $18.13^{+9.30}_{-8.31}$ & $29.26^{+10.81}_{-9.81}$ & $<1.77$ & $2.27^{+1.24}_{-1.02}$ & $18.20^{+9.34}_{-8.34}$ & $29.37^{+10.85}_{-9.85}$ & $<2.45$\\
M23-73488-2 & $0.74^{+0.92}_{-0.70}$ & $4.87^{+7.08}_{-6.10}$ & $8.70^{+8.06}_{-7.08}$ & $<0.88$ & $0.74^{+0.91}_{-0.69}$ & $4.87^{+7.08}_{-6.11}$ & $8.70^{+8.07}_{-7.09}$ & $<0.93$ & $0.71^{+0.88}_{-0.67}$ & $4.89^{+7.11}_{-6.13}$ & $8.74^{+8.10}_{-7.11}$ & $<1.33$\\
M23-77652 & $0.03^{+0.89}_{-0.70}$ & $-1.37^{+7.20}_{-6.34}$ & $-0.79^{+8.12}_{-7.26}$ & $<1.18$ & $0.03^{+0.88}_{-0.69}$ & $-1.37^{+7.20}_{-6.35}$ & $-0.79^{+8.12}_{-7.27}$ & $<1.21$ & $0.03^{+0.85}_{-0.67}$ & $-1.38^{+7.23}_{-6.37}$ & $-0.80^{+8.15}_{-7.29}$ & $<1.53$\\
M23-8083 & $-0.04^{+0.22}_{-0.15}$ & $-0.36^{+1.91}_{-1.62}$ & $-0.37^{+2.03}_{-1.74}$ & $<0.24$ & $-0.04^{+0.22}_{-0.15}$ & $-0.36^{+1.92}_{-1.62}$ & $-0.37^{+2.03}_{-1.74}$ & $<0.25$ & $-0.04^{+0.21}_{-0.14}$ & $-0.36^{+1.92}_{-1.63}$ & $-0.38^{+2.04}_{-1.75}$ & $<0.33$\\
\enddata
%\tablenotetext{1}{}
\tablenotetext{1}{The flux in $0.5-2$ keV derived from $\text{Rate}_\text{soft}$ in Table \ref{tbl:stack}, in $10^{-17}\text{erg s}^{-1}\text{cm}^{-2}$.}
\tablenotetext{2}{The flux in $2-8$ keV derived from $\text{Rate}_\text{hard}$ in Table \ref{tbl:stack}, in $10^{-17}\text{erg s}^{-1}\text{cm}^{-2}$.}
\tablenotetext{3}{The flux in $0.5-8$ keV derived from $\text{Rate}_\text{full}$ in Table \ref{tbl:stack}, in $10^{-17}\text{erg s}^{-1}\text{cm}^{-2}$.}
\tablenotetext{4}{The upper limit of the X-ray luminosity in rest-frame $2-10$ keV, which is the minimum of the upper limits derived from  $\text{Rate}_\text{soft}$, $\text{Rate}_\text{hard}$, and $\text{Rate}_\text{full}$ in Table \ref{tbl:stack}, in $10^{43}\text{erg s}^{-1}$.}
\tablecomments{The entirety of this table is available online at xxx. For fluxes, the values reflect the 16th, 50th, and 84th percentiles of the distributions. The upper limits of luminosities  correspond to a cumulative probability of 0.9987, which is equivalent to $3\sigma$ limits of Gaussian distributions.}
\end{deluxetable*}
\end{longrotatetable}

\section{Results} \label{sec:results}

We now consider the implication of the \textcolor{black}{stacking results. The tentative X-ray detections for the whole sample stack indicate that AGN activity is likely present in LRDs; however, it is still unclear whether these objects have properties similar to other broad-line AGNs.}

\textcolor{black}{One of the} key arguments in support of the hypothesis that AGN are present in LRDs is the observation of broad Balmer, e.g., H$\alpha$, emission lines.
However, type-I AGN activity is not the only plausible source of broad H$\alpha$ lines. Outflows driven by star formation \citep[e.g.,][]{arribas14, davies19,foster20} and supernovae \citep[e.g.,][]{Baldassare16} can also produce broad $\text{H}\alpha$ emission lines. 
%% ACE: can we reference some of the original papers here? I.e. are there some objects with broad Balmer lines that arise from star formation?
To test whether the broad H$\alpha$ lines are indeed linked to AGN activity,
we consider the relation between the X-ray luminosity and the H$\alpha$ luminosity that has been found in low-redshift AGNs \citep[e.g.,][]{ho01,shi10}. Specifically, we adopt the relation measured by \citet{jin12b}, i.e.,
\begin{equation} \label{eq:lxlha}
    \log L_\text{X} [\text{erg s}^{-1}] = 0.83 \times \log L_{\text{H}\alpha} [\text{erg s}^{-1}] + 8.35
\end{equation}
%\begin{equation} \label{eq:lxlha}
%    \log L_{\text{H}\alpha} [\text{erg s}^{-1}] = 1.15 \times \log L_\text{X} [\text{erg s}^{-1}] - 7.66
%\end{equation}
The scatter of this relation is about 0.3 dex. Note that the H$\alpha$ luminosity in this relation corresponds to the broad H$\alpha$ component.
\textcolor{black}{\cite{jin12b} showed that narrow H$\alpha$ lines in AGNs have weaker correlation with X-ray luminosities compared to broad H$\alpha$ lines.}
%The $L_\text{X}-L_{\text{H}\alpha}$ relation allows us to investigate the follow
Assuming that the X-ray emissions and the broad H$\alpha$ lines of LRDs are produced in the same way as the emission in low-redshift type-I AGNs,  
we expect that these objects should follow the observed $L_\text{X}-L_{\text{H}\alpha}$ relation\footnote{We notice that $L_\text{X}$ also tightly correlates with optical and {\oiii} luminosities \citep[e.g.,][]{heckman05, lusso10}. Nevertheless, most LRDs in our sample do not have {\oiii} emission line measurements, and the contribution of the host galaxy to the optical continuum and {\oiii} luminosity is highly uncertain. In contrast, all LRDs in our sample have spectroscopically-confirmed broad H$\alpha$ lines, and thus we focus on the $L_\text{X}-L_{\text{H}\alpha}$ relation.}.

%% ACE: maybe make this a footnote, it sort of disrupts the flow of the section. 
 %, and leave the analysis related to other scaling relations to future studies.

\begin{figure*}
    \centering
    \includegraphics[width=1\linewidth]{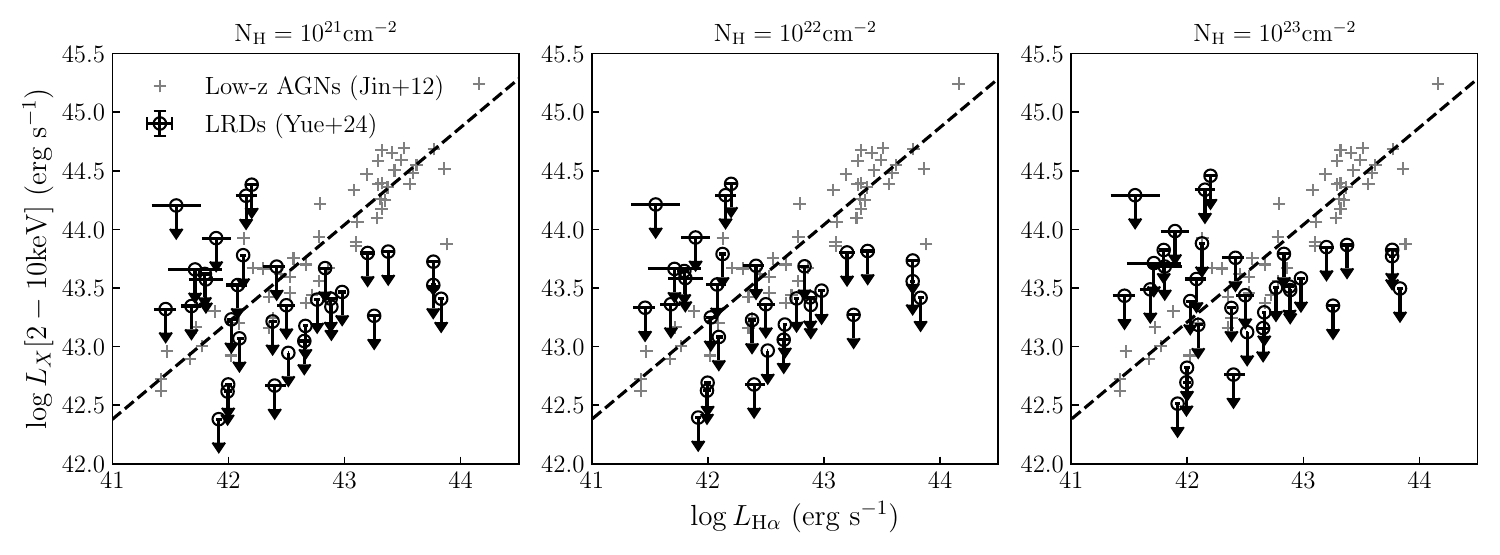}
    \caption{The $L_\text{X}-L_{\text{H}\alpha}$ relation. The gray crosses represent low-redshift type-I AGNs from \citet{jin12}. The dashed line marks the relation as given by Equation \ref{eq:lxlha}. The open circles are the upper limits derived using the soft-band flux limits for individual LRDs. Most of the upper limits are below the $L_\text{X}-L_{\text{H}\alpha}$ relation; in particular, the two LRDs with the largest H$\alpha$ luminosities have X-ray upper limits $\sim1-2$ dex lower than the relation. \textcolor{black}{Column density has very little impact on our results for $N_H\lesssim10^{23}\text{cm}^{-2}$.}}
    \label{fig:LxLHa}
\end{figure*}
%% can you add "(Yue et al., this work)" or something like this to the legend behind "LRDs"

%% ACE: consider adding more sections/ subsection. I think this is really still the results section, the discussion is only later on. And the discussion is really long right now, so I'd at least add some subsections there. 

Figure \ref{fig:LxLHa} shows the $L_\text{X}-L_{\text{H}\alpha}$ relation for low-redshift AGNs and for the LRDs. Most LRDs lie below the $L_\text{X}-L_{\text{H}\alpha}$ relation. In particular, several LRDs with $L_{\text{H}\alpha}\gtrsim10^{43}\text{ erg s}^{-1}$ have X-ray upper limits that are $\sim1-2$ dex weaker than expected from the $L_\text{X}-L_{\text{H}\alpha}$ relation. \textcolor{black}{We notice that the three panels (corresponding to $N_H=10^{21}, 10^{22} \text{ and } 10^{23}\text{cm}^{-2}$) show very little difference, indicating that the impact of column density in our analysis is negligible at $N_H\lesssim10^{23}\text{cm}^{-2}$.} These results indicate that, at least for LRDs with the strongest H$\alpha$ emissions, the X-ray emission is exceptionally weak. A recent study by \citet{wang24} also found similar results, who investigated a luminous LRD at $z=3.1$ and concluded that its X-ray emission must be about 100 times fainter than the expected value given its optical luminosity.

Stacking the whole sample allows us to put a more stringent constraint on the X-ray fluxes of the entire LRD population. Since the LRDs have different redshifts, it is not clear how to convert the stacked photon \textcolor{black}{rates} to physical fluxes and rest-frame luminosities. Instead, we convert the H$\alpha$ luminosities to the expected photon rates for individual LRDs, and average these expected rates using the same weight when stacking (i.e., the exposure time, see Eqn. \ref{eq:rsavg}) to evaluate the expected stacked photon rates. 
%% reference Equation from before

\begin{figure*}
    \centering
    \includegraphics[width=1\linewidth]{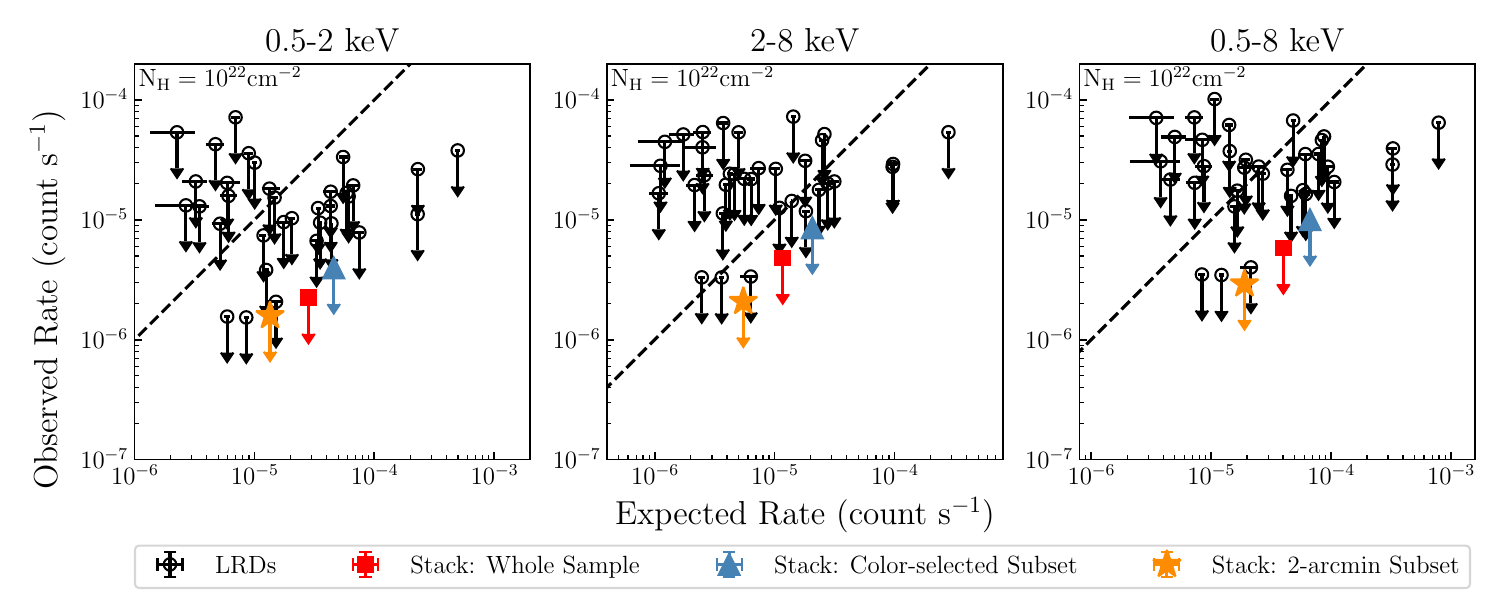}
    \caption{\textcolor{black}{Comparing the observed upper limit of the observed count rates and the expected values from H$\alpha$ luminosities. The dashed lines mark the $x=y$ relation in the plots. The three columns from left to right reflect the soft, the hard, and the full band. 
    %The three rows correspond to different column densities we investigated. 
    The upper limit of the stacked rate in the soft band is $\sim1$ dex below the expected value; this difference is about 0.3 dex for the hard band and 0.8 dex for the full band.
    \textcolor{black}{The conclusions are very similar for the color-selected subset and the 2-arcmin subset, suggesting that the observed X-ray weakness is not a result of selection effects or large off-axis angle observations.} This Figure shows the case for $N_H=10^{22}\text{cm}^{-2}$, and the results are very similar for $N_H=10^{21}\text{cm}^{-2}$ and $N_H=10^{23}\text{cm}^{-2}$.} \textcolor{black}{Note that the whole sample stack has tentative detections (see Table \ref{tbl:stack}), but we still plot the upper limits to be consistent with the subsets that only have non-detections. }}
    %{\em Left:} comparing the soft band stacked rate (the red square) with the expected value from the H$\alpha$ luminosities. The dashed line marks the $x=y$ relation in this plot. The upper limit of the stacked rate is $\sim1$ dex below the expected flux. {\em Right:} same as the left panel, but for the hard band. The stacked flux in the hard band is consistent with the expected value.}
    \label{fig:stackedrelation}
\end{figure*}

\textcolor{black}{
In Figure \ref{fig:stackedrelation}, we present the results for $N_H=10^{22}\text{cm}^{-2}$. We note that column densities have negligible impacts on the results at $N_H\lesssim10^{23}\text{cm}^{-2}$, as demonstrated by Figure \ref{fig:LxLHa} and Table \ref{tbl:fluxlum}}. In the soft band, the upper limit of the \textcolor{black}{whole sample stack} is $\sim1$ dex lower than the expected value derived from the H$\alpha$ luminosity. \textcolor{black}{The differences between the expected stacked rates and the observed upper limits are $\sim0.3$ dex and $\sim 0.8$ dex in the hard and full bands for the whole sample. Stacking the color-selected subset and the 2-arcmin subset leads to similar conclusions. These results indicate again that the  X-ray emission of LRDs is significantly weaker than typical type-I AGNs. }

%However, the stacked photon rate in the hard band is still consistent with the expected level.  %These results %

%Note that all current models of LRDs suggest that the optical attenuation of these objects is large \citep[$A_V\gtrsim2$; e.g.,][]{greene23, pg24}, meaning that the intrinsic H$\alpha$ luminosities are much higher than the observed values. In this case, the discrepancy between the expected X-ray fluxes and the observed upper limits will be even more significant than it appears in Figure \ref{fig:LxLHa}. 

\textcolor{black}{Lastly, we note that recent studies \citep{ananna24, maiolino24} also performed X-ray stacking analysis for LRDs using different samples, who confirmed that LRDs have weak X-ray emission compared to their bolometric luminosities. \textcolor{black}{The weak X-ray emissions} of LRDs is also consistent with \citet{padmanabhan23}, who found that the measured UV emissivity of LRDs would imply an X-ray background  $\sim10$ times higher than constrained by previous observations, assuming LRDs have SEDs similar to typical type-I AGNs.}

\section{Discussion} \label{sec:discussion}

%% ACE: I think here is where the discussion actually starts. 
\textcolor{black}{The tentative X-ray detections after stacking the whole sample indicate that LRDs likely exhibit AGN activity. However, Section \ref{sec:results} shows that the X-ray emissions from LRDs are weak, suggesting that there are some key differences between LRDs and low-redshift type-I AGNs. It is unlikely that the weak X-ray emissions are a result of selection effects or large off-axis angle observations, since the color-selected subset and the 2-arcmin subset has X-ray upper limits much lower than the expected levels. In this Section, we discuss possible explanations for the observed X-ray weakness of LRDs.} 
%The \textcolor{black}{weak X-ray emissions} of LRDs, even after stacking, indicate that  The fact that the K24 subset (which has homogeneous color and compactness cut) and the 2-arcmin sample (which has sharp PSFs in all observations) all give similar results confirm that the weak X-ray emission is not an observational effect. .

%Specifically, we consider models where the rest-frame optical emissions of LRDs are dominated by the obscured and scattered flux from the central type-I AGN. In this case, the H$\alpha$ luminosities measured from the rest-frame optical spectra is only a small fraction of the total H$\alpha$ emission. Therefore, these models imply that the observed $L_\text{X}/L_{\text{H}\alpha}$ ratios of these LRDs should be higher than the intrinsic value. As such, these models are inconsistent with the non-detections of LRDs in the X-ray.

%this result challenges the models where (in which case we expect that LRDs should lie above the $L_\text{X}-L_{\text{H}\alpha}$ relation, as we discussed above). 
%% ACE: did you discuss this?? Remove the parenthesis here. 

%It is also unlikely that the non-detections are results of absorption. 
%% ACE: what do you mean by that? The non-detection of the weak X-ray fluxes?

\subsection{Strong X-ray Absorption}

We first consider the possibility that the faint X-ray emission is a result of absorption.
\textcolor{black}{Our analysis shows that the soft-band upper limit is about 1 dex lower than the expected value even when assuming $N_H=10^{23}\text{cm}^{-2}$.}
%In our model, we assume a hydrogen column density of $N_\text{H}=10^{22}\text{ cm}^{-2}$ for LRDs, which is typical for type-I AGNs. 
%AGNs with higher column density $(N_H\gtrsim10^{23}\text{ cm}^{-2})$ usually have no broad emission lines and appear as type-II or Compton thick AGNs \citep[e.g.,][]{panagiotou19}. 
%Also note that LRDs have redshifts of $z>4$, 
%% ACE: more like z>4
%which means that the soft-band photons have rest-frame energy of $\gtrsim 3-12$ keV and are not sensitive to the photoelectric absorption. 
%To test how this assumption influences our result, we change the column density to $10^{23}\text{ cm}^{-2}$ for the AGN component, which gives an expected soft-band count rate that is 0.9 dex higher than the observed value. 
To move the stacked \textcolor{black}{soft-band upper limit} to the mean $L_\text{X}-L_{\text{H}\alpha}$ relation, we need to adopt a column density of $10^{24.2}\text{ cm}^{-2}$, putting these AGNs into the Compton thick regime. \textcolor{black}{We also note that column densities have little impact on the stacked hard-band rate, which is $\sim0.3$ dex lower than the $L_\text{X}-L_{\text{H}\alpha}$ relation.}
%We also note that $N_H$ has little impact on the expected hard band count rate.
%This hypothesis is unlikely given the presence of broad H$\alpha$ lines. % is in odd with the assumption of type-I AGNs.
%% ACE: absoprtion from what?
%If the weak X-ray fluxes of LRDs are indeed caused by absorption, these objects must have very different absorption mechanisms compared to typical type-I AGNs.
%It is thus unlikely that the X-ray non-detections of LRDs are purely resulted from obscuration.

%Since LRDs in our sample exhibit broad H$\alpha$ lines, 
Most type-I AGNs have $N_\text{H}\sim10^{20}-10^{23}\text{ cm}^{-2}$ \citep[e.g.,][]{koss17,panagiotou19,ananna22a}.
Since the LRDs in our sample exhibit broad H$\alpha$ lines, it is highly unlikely that these objects have such high column densities of $\gtrsim10^{24}\text{ cm}^{-2}$. Some broad absorption line (BAL)
%% define BAL, don't think it has come up before
quasars have high column densities \citep[$N_\text{H}\gtrsim10^{24}\text{ cm}^{-2}$; e.g.,][]{blustin08, rogerson11}, though LRDs with rest-frame UV spectra do not show strong BAL features \citep[e.g.,][]{greene23, maiolino23}\footnote{\textcolor{black}{We note that about 10\% LRDs exhibit strong H$\alpha$ absorptions \citep[e.g.,][]{matthee24, maiolino24}, but these absorption lines are narrow and distinct from BAL features}}.
We also notice that some low-redshift type-I AGNs have $N_\text{H}>10^{24}\text{cm}^{-2}$ \citep[e.g.,][]{ananna22b}; however, these AGNs are very rare, which only occupy a small fraction $(\lesssim5\%)$ of the entire type-I AGN population. In comparison, LRDs are numerous, the number density of which is $\sim1$ dex higher than the quasar luminosity function and is only $\sim2$ dex lower than the galaxy luminosity function at similar redshifts \citep{greene23,kocevski24}. We thus consider it unlikely that absorption is the only reason for the low X-ray fluxes of LRDs.
%
 
%% obscuration?
%Based on the above considerations, we suggest that it is unlikely for absorption to be the only reason for the low X-ray fluxes of LRDs. 

\subsection{Intrinsically faint X-ray emission}

Alternatively, we suggest that LRDs might be a population of AGNs with intrinsically-faint X-ray emission, implying a low $L_\text{X}/L_{\text{H}\alpha}$ ratio. 
%Observations of low-redshift AGNs have shown that AGNs with high Eddington ratios are likely to have weak X-ray emissions. 
%The ratio between optical and X-ray flux for AGNs (i.e., $\alpha_\text{OX}$)
%We notice that the ratio between optical and X-ray flux (i.e., $\alpha_\text{OX}$) increases with Eddington ratios. \citet{matthee24} have estimated the Eddington ratios of LRDs to be $\lambda_\text{Edd}\sim0.07-0.4$
Specifically, the inner accretion disk and the corona of LRDs might be different from other type-I AGNs. 
It is also possible that LRDs have unique broad line region properties, such that their H$\alpha$ line fluxes are higher than other type-I AGNs.

Previous studies have pointed out that the ratio between optical and X-ray fluxes for AGNs (i.e., $\alpha_\text{OX}$) increases with Eddington ratio \citep[e.g.,][]{vasudevan07,jin12c}. The Eddington ratios of LRDs have been estimated to be $\lambda_\text{Edd}\sim0.05-2$ \citep[e.g.,][]{harikane23, maiolino23}, which is comparable to the AGN sample used to derive the $L_\text{X}-L_{\text{H}\alpha}$ relation in \citet{jin12b}. 
%We notice that the two LRDs in our sample with the most stringent upper limits on $L_\text{X}/L_{\text{H}\alpha}$ (CEERS\_00746 and CEERS\_00672) have $\lambda_\text{edd}=1.47$ and 0.65, respectively
%We note that 
%. The intrinsic Eddington ratio of LRDs might be higher than the reported values due to attenuation \citep[e.g.,][]{matthee24}. We note that 
%
However, we note that the intrinsic Eddington ratio of LRDs might be higher than the reported values due to dust attenuation \citep[e.g.,][]{matthee24}. 
Therefore, high Eddington ratios might contribute to the low observed $L_\text{X}/L_{\text{H}\alpha}$ ratios of LRDs.
Recent theoretical models have shown that super-Eddington accretion could indeed be a possible solution to explain the X-ray weakness of LRDs \textcolor{black}{\citep[e.g., ][]{pacucci24,volonteri24}.}
Since the estimated Eddington ratios of LRDs, however, still have large uncertainties, we leave a more quantitative investigation to future studies.

%Therefore, Eddington ratios might not be the main driver that produces the weak X-ray emissions of LRDs.

We note that the intrinsic $L_\text{X}/L_{\text{H}\alpha}$ ratios of LRDs might be even lower than indicated in Figure \ref{fig:LxLHa}. Most current modes of LRDs suggest large dust attenuations \citep[$A_V\gtrsim2$; e.g.,][]{pg24, kocevski24}. Furthermore, some models have suggested that the broad emission lines of LRDs are produced by scattered light from the central type-I AGN, which is only a small fraction of the intrinsic emissions \citep[e.g.,][]{greene23}. Consequently, the intrinsic H$\alpha$ luminosities of LRDs might be larger than the observed values, leading to lower intrinsic $L_\text{X}/L_{\text{H}\alpha}$ ratios.

\subsection{Broad Balmer emission lines due to outflows}

%Another plausible scenario 
\textcolor{black}{Another scenario previously proposed is that}
%is that 
%LRDs are indeed not AGNs or only have very weak AGN activity. In this case, 
AGN activity \textcolor{black}{only has partial} contribution to the observed broad H$\alpha$ lines of LRDs. Some other mechanisms \textcolor{black}{can also} produce H$\alpha$ emission lines with FWHM$\gtrsim1000\text{ km s}^{-1}$, including strong outflows driven by star formation or supernovae \citep[e.g.,][]{Baldassare16,davies19,foster20}. 
%\textcolor{black}{}
%% ACE: add references here
This hypothesis agrees with the result from recent MIRI observations \citep[][]{pg24}, who found that starburst galaxies might have a major contribution to the near-to-mid infrared SED of LRDs. \textcolor{black}{However, recent NIRSpec observations have shown that LRDs have narrow {\oiii} emission lines \citep[e.g.,][]{maiolino24}, which are hard to explain by the outflow scenario. Therefore, we consider it unlikely that outflows have major contributions to the broad H$\alpha$ emission linesm in LRDs.} 

Future {\em JWST}/NIRSpec observations will characterize the fluxes, kinematics, and spatial extents of the UV and optical emission lines \textcolor{black}{for larger LRD samples}, which will provide more clues about the contribution of outflows to the broad H$\alpha$ lines.

%We emphasize that 

% The X-ray non-detections, together with the observed features in other wavelengths, suggest that LRDs are intrinsically different from typical type-I AGNs.
In any case, our results indicate that we need to be cautious when applying previous knowledge about type-I AGNs to these LRDs. In particular, the bolometric luminosities and the SMBH masses of LRDs estimated from the scaling relations for type-I AGNs might have significant systematic uncertainties. 

%Finally, we emphasize that our result does not rule out the AGN hypothesis of LRDs, the nature of which is still highly uncertain. So far, it is still challenging to find a model that explains all the observed features of LRDs, including the V-shaped SED in rest-frame optical \citep[i.e., a blue slope at $\lambda_\text{rest}\lesssim0.5\mu$m and a red slope at $\lambda_\text{rest}\gtrsim0.5\mu$m; e.g.,][]{kocevski23, greene23, pg24}, the weak mid-infrared emission \citep[e.g., ][]{pg24, wang24}, and the X-ray non-detection described in this Letter. Future multi-wavelength follow-up studies will provide more clues about this mysterious population of objects.

\section{Conclusions} \label{sec:conclusions}

We investigate the X-ray properties of 34 LRDs using a stacking analysis. The LRD sample is compiled from the literature and is selected to show broad H$\alpha$ emission lines with FWHM $>1000\text{ km s}^{-1}$, indicating possible type-I AGN activity. \textcolor{black}{None of the individual LRDs are detected in soft or hard bands. After stacking the whole sample, the soft, hard, and full bands exhibit tentative detections with $2.9\sigma$, $3.2\sigma$, and $4.1\sigma$ significance, respectively. These tentative detections indicate that AGN activity likely presents in LRDs.} 

\textcolor{black}{We further compared the X-ray flux upper limits of LRDs to their H$\alpha$ luminosities.}
\textcolor{black}{The upper limit on the soft (hard) band flux is $\sim1$ dex  ($\sim0.3$ dex) lower than the expected level from the $L_\text{X}-L_{\text{H}\alpha}$ relation of low-redshift type-I AGN,} \textcolor{black}{suggesting that LRDs are weak in X-ray emissions. This conclusion holds for the subset with uniform color cuts and the subset with small off-axis angle observations.}%, while the upper limit on the hard band flux is consistent with the relation. 
%% only in the soft band, not in the hard band, I'd mention that here

Our result suggests that LRDs might be a population of AGNs with distinct properties compared to \textcolor{black}{previously-identified type-I AGNs, i.e., LRDs might have intrinsically weak X-ray emissions.
 We find it difficult to explain the observed weak X-ray fluxes of LRDs solely by absorption, and it is also unlikely that fast outflows have a major contribution to the broad H$\alpha$ lines. }
%Alternatively, we discuss two plausible hypotheses to explain the observed low $L_\text{X}/L_{\text{H}\alpha}$ ratios: 
%(1) LRDs have strong X-ray absorption, which we consider unlikely; %2) more plausible scenarios... 
%(1) ; 
%(2) outflows have a major contribution to the observed broad H$\alpha$ emission lines. 
In any case, caution should be taken when applying the empirical relations derived from other type-I AGNs to the LRD population.

The sample of LRDs is rapidly increasing thanks to the ongoing imaging and spectroscopic surveys with {\em JWST}. With a larger LRD sample in the near future, we expect that a stacking analysis will put even stronger constraints on the X-ray properties of LRDs. Furthermore, upcoming {\em JWST}/NIRSpec observations will \textcolor{black}{greatly increase the sample of spectroscopically-confirmed LRDs and will} reveal the fluxes, kinematics, and spatial extents of the other optical and UV emission lines of these objects. % shedding more light on the nature of this population. 
Future X-ray missions, such as the Advanced X-Ray Imaging Satellite \citep[AXIS;][]{axis}, will offer much deeper images than {\em Chandra} and will likely solve the puzzles about the X-ray properties of LRDs.
%Other observations, like spatially-resolved emission line diagnostics by NIRSpec/IFU, will also provide critical clues about the nature of LRDs.
%% Let's add something about AXIS solving all of this in the future... it'll make Erin happy :)

%% ACE: why do we want to re-do this with O3? What's the advantage of O3 to Halpha?

\begin{acknowledgments}
\textcolor{black}{We thank the anonymous referee for valuable comments.}
We thank the valuable comments from Thomas Connor.
TTA acknowledges support from ADAP grant 80NSSC23K0557. 
TM and the development of the CSTACK tool are supported by the UNAM-DGAPA PAPIIT IN114423.
\textcolor{black}{The scientific results reported in this article
are based on observations made by the Chandra X-ray
Observatory. This work made use of software provided by
the Chandra X-ray Center (CXC) in the CIAO application
package.}
%This paper employs a list of Chandra datasets, obtained by the Chandra X-ray Observatory, contained in~\dataset[DOI: X]{https://doi.org/X}.
\end{acknowledgments}


\vspace{5mm}
\facilities{Chandra(ACIS-I)}


\software{CSTACK, CIAO, sherpa\citep{sherpa,sherpa2}, xspec\citep{xspec}}


\bibliography{sample631}{}
\bibliographystyle{aasjournal}



\end{document}